\documentclass[]{spie}  

\usepackage{amsmath,amsfonts,amssymb}
\usepackage{graphicx}

\usepackage[colorlinks=true, allcolors=blue]{hyperref}

\title{POKEMON: print optimization for kilo-fiber experiments using micro-optics and nanostructures}

\author[a,b]{Megan Delamer}
\affil[a]{Pennsylvania State University}
\affil[b]{Center for Exoplanets and Habitable Worlds}
\author[a,b]{Suvrath Mahadevan}
\author[c]{Chad Bender}
\author[a,b]{Ceiwynn Longworth}
\author[c]{Roger Angel}
\affil[c]{University of Arizona}

\author[c]{Joel Berkson}
\author[c]{On To Sonja Choi}
\author[a]{Kathleen Gehoski}
\author[c]{Andy Monson}
\author[d]{Christian Schwab}
\affil[d]{Macquarie University}

\authorinfo{Further author information: (Send correspondence to M.D.)\\M.D.: E-mail: mmd6393@psu.edu}

\begin{document} 
\maketitle

\begin{abstract}
~~~The most pressing problems in modern astrophysics have often required the largest telescopes. With the cost scaling of mirror diameters, the field as a whole is faced with a challenge- how to replicate or improve on the collecting area and sensitivity of the current generation of ELTs, which already boast 30\,m class apertures and are multi-billion dollar facilities.  One such approach is being pursued by the  Large Fiber Array Spectroscopic Telescope (LFAST) - a scalable array telescope. Each element of the array will consist of multiple mirrors each feeding to an individual fiber; with those fiber feeds feeding optical and infrared spectrometers. Coupling fiber bundle to spectrometer slit input must be optimized to take full advantage of the photon collecting ability of the telescope array, requiring precise alignment of microlenses to each fiber. Advances in two photon polymerization processes (2PP)  now allow for optical quality microlenses with wavefront aberrations as small as $\lambda/20$\cite{Imiolczyk_ultra_2024} to be created, opening up the design parameter of bespoke optical design and custom fabricated lenses. We present our approach to tackling these coupling problems with rapid prototyping and detailed quantification of the tolerances of the lenses. Our approach leverages our access to the Nanoscribe GT2 system at Penn State, enabling tests of new optically transparent resins like IPX-Clear to explore multiple design approaches. Our goal is to share our results and enable wider use of these techniques for astronomical applications. 

\end{abstract}


\section{INTRODUCTION}
\label{sec:intro}  
~~~Several subfields of astrophysics are ``photon-starved", where the discovery space is limited by the number of photons that can be collected from a particular source. While advancements in detector technology significantly improved amplitude of signals throughout the 20th century, the biggest gains now are realized as the primary mirror grows in diameter. The newest generation of ground-based observatories are often referred to as the Extremely Large Telescopes (ELTs) and are projected to come online in the upcoming decades. While every one of these facilities will be extremely sensitive due to their 30 meter class mirrors and be capable of exploring the fundamental physical laws that drive our universe, the cost of such facilities will also be astronomical. Telescope costs scale non-linearly with the diameter of the primary mirror (D); first order approximations give cost as proportional to D$^{n}$, where $n$ ranges from 2.0-2.8 depending on design choices and decade of construction \cite{Meinel_overview_1978,vanbelle_scaling_2004}. Some exceptions to this relationship were able to reduce costs while sacrificing operational flexibility; fixing the altitude of the telescope (e.g. the Hobby-Eberly Telescope\cite{ramsey_early_1998,hill_hetdex_2021}) reduces the amount of structure needed to support the mirrors, though it also reduces the number of visible targets at any given time.

An alternative to the traditional single aperture construction is the combination of many smaller apertures feeding light into a single instrument, often using optical fibers to connect the individual apertures as first proposed by Angel et al.\cite{Angel_very_1977}. The Miniature Exoplanet Radial Velocity Array (MINERVA)\cite{swift_minerva_2015} and MINERVA-Australis\cite{addison_minera-australis_2019} both utilize a number of smaller aperture telescopes that can either work independently or in concert depending on the science case. The approaches used to combine light from the individual apertures limit the number that can be utilized in these designs for various reasons and none of them have the collecting area of an ELT mirror. The major advantage in multi-aperture designs is the lowered cost of support structures; although glass is expensive and a multi-aperture set-up does not decrease that expense, the cost of both domes and the physical mounts for fully steerable telescopes is approximately proportional to D$^3$\cite{vanbelle_scaling_2004}. Upcoming projects such as Multi Array of Combined Telescopes (MARCOT)\cite{Roth_MARCOT_2022} and PolyOculus\cite{Eikenberry_polyoculus_2019} are taking advantage of this proportionality to provide a low cost alternative to traditional monolithic mirrors, but still at smaller effective aperture sizes than the ELTs.

LFAST is new telescope concept designed to scale to a collecting area equivalent to that of the ELTs, but able to be constructed at a very low cost\cite{Angel_LFAST_2022}. The array is designed using 0.76\,m primary mirror apertures mounted in groups of 20, which provides the equivalent collecting area of a traditional 3.5\,m aperture for each system. The full telescope array will consist of 132 20-unit systems for a total collecting area of 1,200\,m$^2$\cite{Angel_LFAST_2022,Bender_LFAST_2022}. To save further on overall cost, the systems will not be housed in domes, instead they will be provided with automated removable covers and stored pointing 15$^o$ below the horizon to avoid dust settling on the mirror surfaces\cite{Young_LFAST_2022}. Each individual aperture will feed to its own optical fiber with an 18\,micron diameter core to have an on sky footprint of 1.4\,arcsec\cite{Berkson_LFAST_2022}. The light from these fibers will be combined into one rectangular fiber and fed into one or more spectrometers. A major technical challenge is combining the light with minimal losses, such that the system is able to take full advantage of the photon collecting ability of each individual mirror. LFAST fibers will be combined using a two dimensional microlens array, with each microlens placed precisely over the center of the optical fibers to minimize photon loss\cite{Angel_LFAST_2022}.

While standard fabrication techniques for creating optical elements are capable of producing spherical lenses down to 0.3\,mm in diameter, complex systems such as the microlens array needed to combine the light from 2640 individual telescopes are more challenging. Instead, we are creating the microlenses using 2PP, a 3D printing method that utilizes a femtosecond laser and a high numerical aperture focusing objective to transition a pool of liquid resin into a solid polymerized state\cite{goppert_uber_1931,maruo_three_1997}. The advantage of using two photons in quick succession to bridge an energy gap rather than using a single photon of higher energy is the single photon method will induce polymerization all along the path of the laser, while the cross section for the two photon method produces a very small region (the voxel) where sufficient energy is available to polymerize resin\cite{fischer_three_2013}. Recently available resins with high transmission across the optical and near IR wavelength range, such as IPX-Clear\footnote{https://www.nanoscribe.com/en/products/ipx-photoresins/photoresin-ipx-clear-for-highly-transparent-microoptics/}, are ideal for the planned wavelength coverage of LFAST (390\,nm - 1700\,nm)\cite{Angel_LFAST_2022,Bender_LFAST_2022}. In this work, we present the ongoing work to produce a 2x10 microlens array for the initial LFAST prototype currently under construction at the University of Arizona. We will describe our fabrication process, the current status of the prototype, and future work.

\section{DESIGN AND FABRICATION}
\subsection{Design}
~~~The current prototype consists of 2 x 10 array of spherical microlenses with a pitch of 250\,microns; each lens has a radius of curvature (ROC) of 400\,microns and a diameter of 245\,microns, giving a sag of 19.2\,microns. The current design features circular lenses, but the nature of 3D printing with 2PP would allow for a straightforward change to further improve the fill factor of the array\cite{Angel_LFAST_2022}. The design will allow each lens to image the 18\,micron fiber core to a single rectangular fiber that will feed one or more spectrometers\cite{Bender_LFAST_2022}.
\subsection{Fabrication}
\begin{figure}
    \centering
    \includegraphics[width=0.95\linewidth]{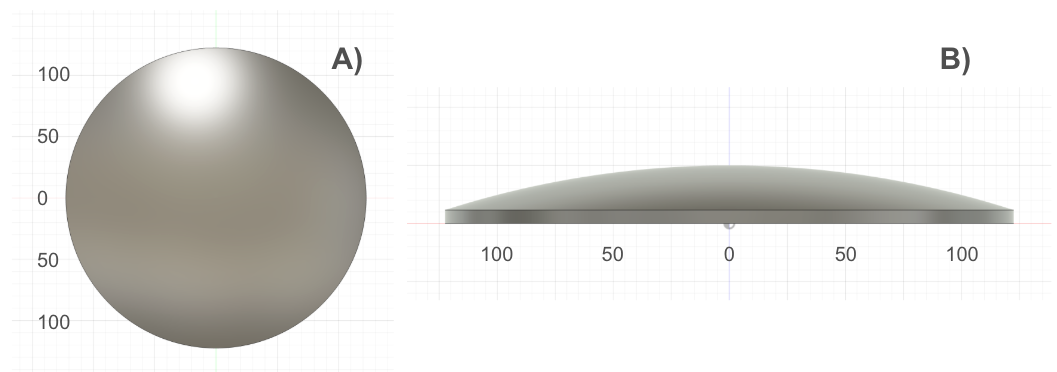}
    \caption{A) Top down view of the lens model, with gridlines indicating 10\,micron distances. B) Side view of the lens model, with gridlines indicating 5\,micron distances.}
    \label{fig:model}
\end{figure}

~~~We produced a computer aided design (CAD) model of a single lens using the free Autodesk Fusion software, which was exported in STL --- a file structure containing only the surface geometry of a 3D object --- format for 3D printing. A base was added to the design to promote better adhesion to the substrate and reduce the possibility for shrinkage to pull the edge of the lens away from the substrate, which brings the total designed height to 25\,microns. \autoref{fig:model} shows the lens model, including this added base.

The nanofabrication facility in the Materials Research Institute at Pennsylvania State University has a Nanoscribe Photonic Professional printer, which uses a 780\,nm femtosecond laser with a pulse duration of 80-100\,fs and repetition rate of 80\,MHz as the light source for triggering polymerization in resists. The laser is focused into the resin with a Carl Zeiss Axio Observer inverted microscope, which also enables monitoring of the printing process. Three focusing objectives can be used with the system, each trading off between reachable printing area and resolution. The 25x objective with 0.8\,NA best suited the lens design, providing relatively small features while allowing for the entire lens to be printed within the lateral scanning range of the laser, eliminating the reduction in optical quality that would come from combining multiple stage movements together or the substantial increase in print time moving only the piezo system would cause. Due to the small dimensions of the design, the lenses were printed onto a 25\,mm x 25\,mm x 0.7\,mm fused silica substrate which enables easier characterization of the lenses and is accounted for in the LFAST optical design for the fiber output feed.

\begin{figure}
    \centering
    \includegraphics[width=0.95\linewidth]{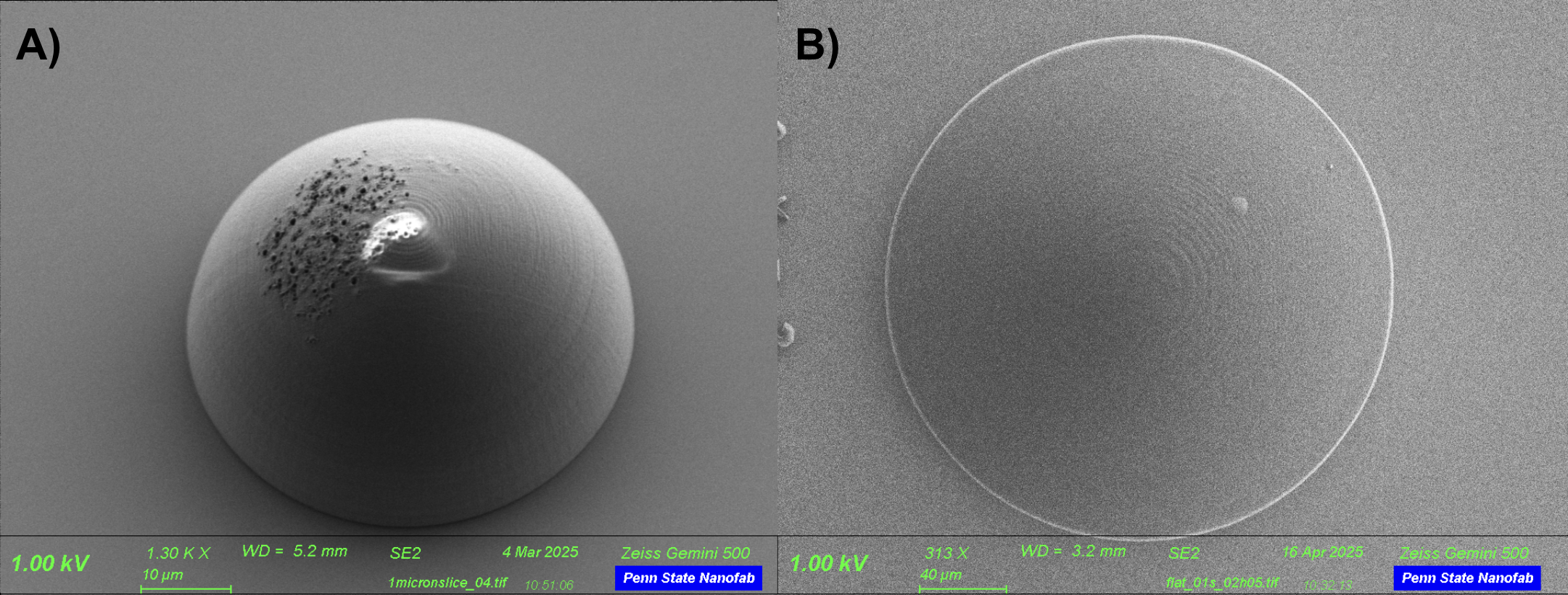}
    \caption{SEM images taken with the Zeiss Gemini 500 FESEM in the nanofabrication facility at the Pennsylvania State University Materials Research Institute. A) Image of a 50\,micron diameter spherical half-ball lens printed for initial testing with a pitted surface due to overexposing and burning the resin. B) Image of the 245\,micron lenses with a bubble in the upper righthand quadrant.}
    \label{fig:sem-images}
\end{figure}

The Photonic Professional utilizes proprietary DeScribe software to convert STL files into print jobs readable by the machine (generalized writing language (GWL) files). The import wizard offers significant flexibility in how jobs are split; there is the vertical distance between layers (slicing) and the distance between parallel lines within each layer (hatching). Arbitrarily larger or small distances can be chosen for the slicing and hatching, but the size of the voxel sets realistic limits on both of these choices. Voxel size is given in two dimensions, length $l$ and diameter $d$, and is dependent on the laser, the objective, and the selected resin. The approximate dimensions of the voxel in the Photonic Professional with the 25x objective are $l=3.3$\,microns and $d=0.6$\,microns. For a smooth, closed surface, the chosen slice and hatch can be no larger than these values or adjacent lines will not overlap and unpolymerized material will leak out during the develop phase. However, optical quality surfaces necessitate smaller choices because large slicing steps will be visible on any curved surface. Selecting too small of a slice or hatch introduces the possibility of overexposing the resin, which can produce bubbles or even burn the resin (see \autoref{fig:sem-images} for examples).

Determining the correct amount of power to deliver such that small slices can be chosen to create optical quality structures, adequate polymerization will ensue to create a solid surface, and prevent overdosing is a challenging, iterative process because of the many dependencies involved. A simple expression for the dose delivered while printing is:
\begin{equation}
    D = \frac{LP^n}{v}
\end{equation}
where $LP$ is the laser power, $v$ is the write speed, and $n$ is an exponent related to nonlinear absorption that ranges from 2-4 depending on the resin. However, this expression fails to include that the size of the voxel, therefore degree of overlap between slices, is directly dependent on the laser power. It also fails to capture the complexity of the polymerization process, which is influenced by temperature, humidity, and is time-dependent. For larger structures, the time dependency has only a minor effect because the polymerization process terminates before the laser has swept back to print the next layer, but the peak of a lens has a number of small, rapidly printed layers where polymerization is less likely to have ceased before the next layer has begun.

An alternative to a solid structure is printing with the shell and scaffold method, which offers significant improvements in the speed of a print. In this method, the slice and hatch apply only to a thin shell covering an internal structure composed of either planar or triangular scaffolds. This method can leave pockets of resin unpolymerized within a structure; UV curing can solidify those pockets, but only in resins formulated to be UV sensitive.

\subsection{IPX-Clear}
~~~IPX-Clear is a new 2PP resin formulated by Nanoscribe specifically for micro-optics and optimized for usage with the Nanoscribe Quantum X printer, but compatible with the Photonic Professional as well. While the polymerized resin must be UV cured for optimal clarity, the transmittance from 350\,nm to 1550\,nm is $>$95\% and the lowest transmission within the LFAST wavelength regime is still $>88$\% for a 1\,mm thick sample, which is far thicker than our lens design. Other available resins marketed for micro-optics and specifically designed for the Photonic Professional are limited to $\sim$90\% transmittance and have significant decreases in transmission around 500\,nm that would curtail the full range of LFAST. 

\subsection{Fabrication Workflow}
~~~The workflow for fabrication with the Photonic Professional follows a few key steps, with deviations made only in the selection of print parameters to ensure process control. The fused silica substrates are rinsed in acetone, then IPA for $\sim$30\,s and blown dry with nitrogen. The substrate is then plasma etched  using a SAMCO AQ 2000 with O$_2$ for 5\,minutes to promote better adhesion of the lenses to the substrate. A pool of IPX-Clear is deposited onto the substrate and allowed to sit for 10\,minutes prior to beginning the print to allow all components to thermalize. After the print but prior to the develop step, the substrate is baked on a hot plate at 65$^o$, then 95$^o$, then 65$^o$ again each step lasting 2\,minutes. This step serves a dual purpose of further promoting adhesion and allowing polymerization to continue for a short time after the print. After the bake, the print is placed in a develop bath (mr-dev~600) for 20\,minutes to remove all unpolymerized material, followed by a 2\,minute IPA bath to remove any remaining mr~dev~600. Once dried, the lenses are cured under a UV light source for 10\,minutes.

\section{CHARACTERIZATION}
\label{sec:characterization}
~~~There are three main areas of concern regarding the creation of a microlens array that will maximize the photon collecting area of the LFAST prototype: 1) The profile of the lenses must adhere to the design in order to ensure they focus at the right distance with low wavefront error, 2) the surface roughness of the lenses must be low, and 3) the printing process must result in identical lenses at precise spacings in order to ensure they are directly over the fiber cores and maximize the light captured from each individual fiber. Design choices can introduce additional concerns, e.g. utilizing the shell and scaffold method may leave pockets of resin with a lower polymerization fraction compared to that of a solidly printed lens, altering the optical pathway encountered by photons. There is also a minor concern regarding how well the lenses adhere to the substrate, other work has seen the edges of printed structures peel slightly upwards, and the microlens array must be robust enough to be shipped and handled.

 Due to shrinkage in the printed material, which can vary depending on the laser dose delivered, the printed lenses do not perfectly match the design and this effect can be highly anisotropic. An iterative design process can be used to compensate for this change, but the shape deviations must be measured precisely. While SEM images can reveal surface roughness and burns caused by overdosing, determining how closely the printed lens adheres to the model requires optical profilometry. The Zygo Nexview 3D Optical Surface Profiler uses white light interferometry to determine the surface contours of a sample; the vertical resolution of the instrument is on the order of a few angstroms, more than sufficient to detect profile differences and surface roughness at the level with which we are concerned. The testing is also non-contact, causing no damage to the lenses, allowing for precise knowledge of the actual prototypes that will be incorporated into the LFAST array and not simply reproductions that should be identical. All data taken with this instrument was analyzed using Mx, a proprietary software package that calculates surface roughness and maximal deviations in z from a programmed model. As the sharp angle at the edge of the lenses is too steep to be captured by the profilometer, analysis takes place on the inner 235\,microns of all lenses to give a 5\,micron leeway.

\begin{figure}
     \centering
     \includegraphics[width=0.95\linewidth]{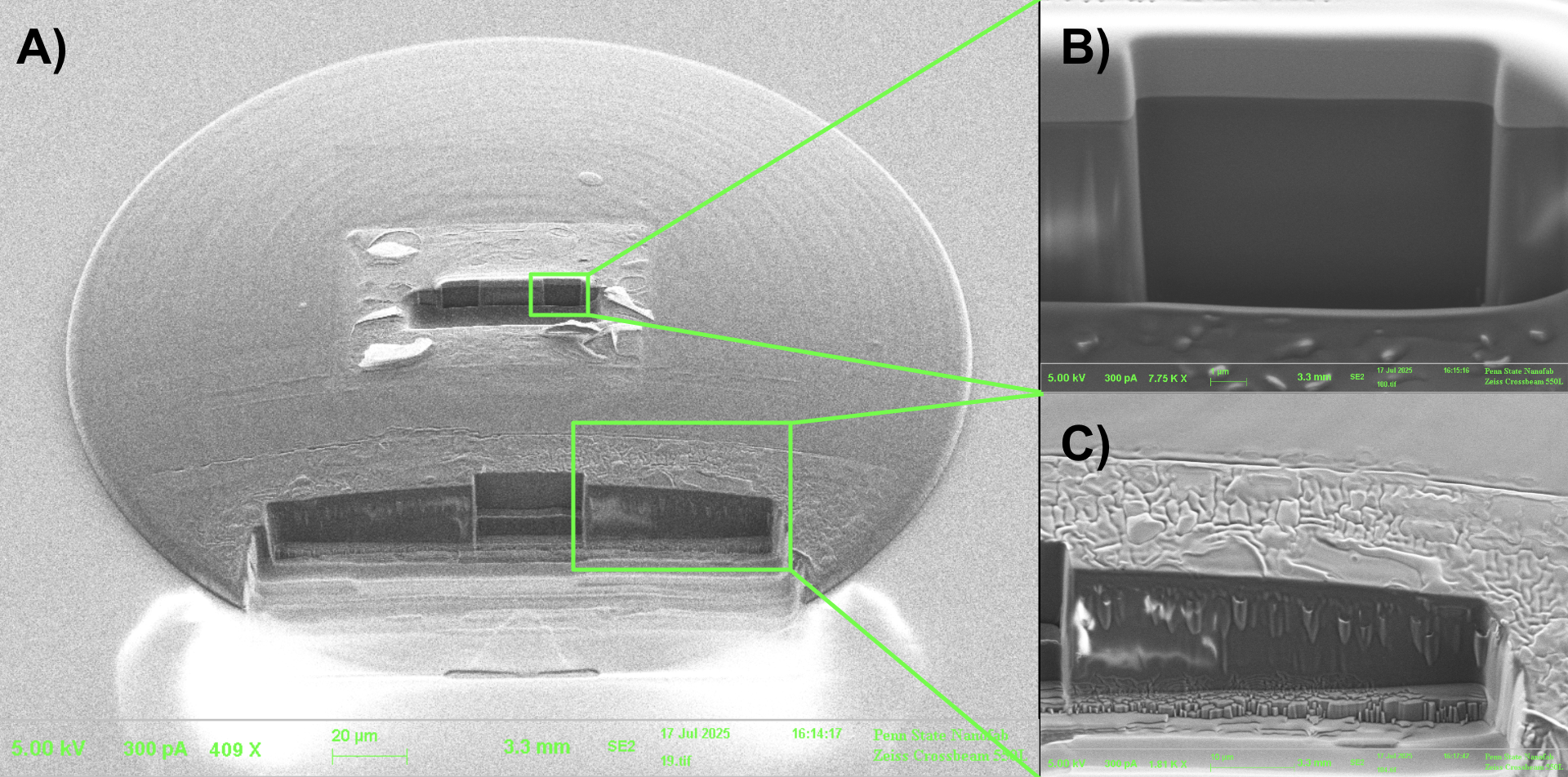}
     \caption{Images of a solid lens taken post removing material with FIB in the Zeiss Crossbeam 550L FESEM. A) An image of the full lens after removing material from two different regions. At the peak of the lens, a carbon layer was deposited locally to prevent the damage that can be caused by the ions. Near the base of the lens, a carbon layer was not applied and significant damage was caused to the sample. B) An inset of the cut made in the protected region, revealing a smooth interior. C) An inset of the cut made in the unprotected region, where ions have damaged the surface and caused scouring marks through the lens material.}
     \label{fig:fib}
 \end{figure}

 Focused ion beam (FIB) is a technique used in concert with SEM imaging to look at the internal structure of materials. In this technique, a narrow beam of ions is used to sputter away material in a region typically a few nanometers across. FIB can cause significant damage to a sample (\autoref{fig:fib}, reducing the potential for truly understanding the internal structure. A conductive layer of carbon applied as a coating significantly reduces the damage to the sample, with FIB conducted on protected areas revealing that the interiors of the lenses printed with the solid method do not show any residual slicing and hatching structure (\autoref{fig:fib}).

 \subsection{Testing Stepping Precision}
 ~~~The Nanoscribe Photonic Professional has two methods of navigating the X-Y plane: moving the stage and moving the piezo. Moving with only the piezo ensures a precise step, with movement error $<$10\,nm, but the piezo has an extremely limited range and can only step 300\,microns in X and Y before needing to be reset. In contrast, the stage can move multiple millimeters, but does so with an error in step precision of $\sim$1\,micron. We aim to understand how the error in stepping precision compounds with multiple stage movements to test if moving via the stage (the default movement mode of the printer) is sufficient for our needed precision. To test this, we printed a 10x10 grid of crosshairs and used optical profilometry to determine the distances between individual points as well as the distance between the first and last step in a column. The programmed step between each crosshair was 20\,microns and the actual step averaged 19.9\,microns in both the X and Y directions. The errors did compound slightly as the total distance between first and last points averaged 179.7\,microns rather than the programmed 180\,microns. Additional tests with longer steps are needed to fully determine if the stage movement will be acceptable for the microlens array, but these early results are promising.

\section{CHALLENGES IN DEVELOPMENT}
~~~The greatest challenge in development to this point has been determining the optimal laser dose to induce polymerization evenly across the lens and cause any shrinkage to be isotropic. As IPX-Clear was developed to be used in the creation of micro-optics with the Quantum X printer, the only recipes available are optimized for two-photon grayscale lithography (2GL). This is a 2.5D printing method in which the printer can actively adjust the laser power to change the voxel size only at the surface of the lens, while using the full polymerization volume internally to improve the printing speed. The Photonic Professional does not offer this same level of flexibility through the DeScribe STL model import method.

\begin{figure}
    \centering
    \includegraphics[width=0.95\linewidth]{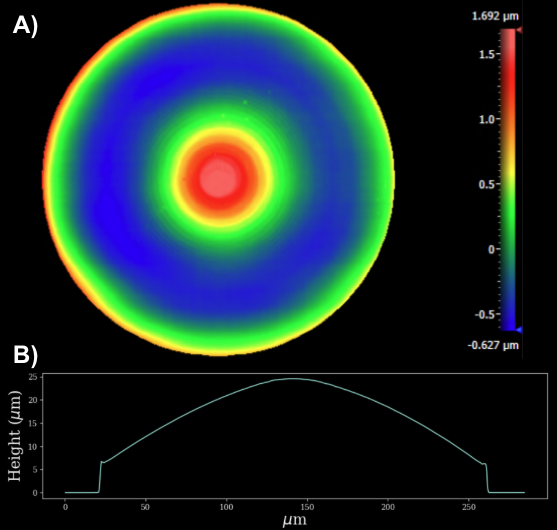}
    \caption{A) Optical profilometry data of a lens printed using the suggested shell parameters from the 2GL print method with a best fitting spherical model removed. B) The height of the lens as a function of distance along a slice running parallel to the z-axis.}
    \label{fig:profile}
\end{figure}

For the initial attempts at creating lenses, we took the suggested parameters for the outer shell of the 2GL methodology and altered them to better suit the Photonic Professional printer (see \autoref{tab:printsets} for exact parameters). As can be seen in \autoref{fig:profile}, this resulted in a lens with a noticeably peaked profile, deviating from a best fitting spherical model by nearly 2.5\,microns. The innermost region of lenses printed with these parameters does have a surface roughness of only 30\,nm, a reasonable level for LFAST, but the profile deviation is well outside an acceptable range.

\begin{table}[h!]
    \centering
    \begin{tabular}{c|c|c|c}
       Name  & 2GL Shell like & 2GL Interior like & IP-S like \\
       \hline
        Slice (microns) & 0.1 & 1.2 & 0.2\\
        Hatch (microns) & 0.1 & 0.25 & 0.2\\
        Hatching Angle Offset ($^o$) & 137.508 & 137.508 & 137.508 \\
        Laser Power (mW) & 17.5 & 17.5 & 17.5 \\
        Write Speed  &40000 & 40000 & 40000\\
    \end{tabular}
    \caption{Print sets tested for solidly printed microlenses, adapted from recipes for similar materials and processes.}
    \label{tab:printsets}
\end{table}

Another adapted recipe was based on the interior portion of 2GL lens; this featured 1.2\,micron slices and 0.25\,micron hatches. This resulted in an extremely low surface quality, but a lens profile that was spherical, with isotropic shrinkage bringing the best-fit ROC to 375\,microns. As this test was conducted identically to the previous one, with the only change coming in the choice of slice and hatch, we determined that the uneven shrinkage in the first print was due to dose accumulation at the peak of the lenses, where small diameter slices are more printed more rapidly. To combat the dose accumulation, we tested increasing the waiting duration between successive slices using the \texttt{PiezoSettlingTime} command in DeScribe. The test included 5, 10, 15, 20, and 30\,second wait times and the increase in time provided a reduction in how much the lens deviated from a spherical model until the lowest deviation was $<0.9$\,microns. Unfortunately, this deviation was still above an acceptable threshold and caused two other problems: 1) every increase in wait time decreased the overall height of the lens, with the 30\,s settling time lens being nearly 6\,microns shorter than the design and 2) the print time for an individual lens exceeded 2\,hours, making this an impractical choice for the final array which will require 2640 microlenses.

Previous micro-optics produced with the Photonic Professional printer have used IP-S as the resin with the same 25x objective we are using with IPX-Clear. We switched to using slice and hatch parameters from an IP-S micro-optics recipe (see \autoref{tab:printsets}) and conducted tests with an array of writing speeds and laser powers. A laser power of 7.5\,mW reduced the deviation from spherical to $<1$\,micron, but resulted in a consistently shorter height than the planned 25\,microns, indicating material in the base was not fully polymerizing. An additional issue with a laser power this low was a lack of consistency in lenses printed on the same substrate; heights of paired lenses differed by $>1$\,micron, despite undergoing identical processing on the same substrate.

\begin{figure}
    \centering
    \includegraphics[width=0.95\linewidth]{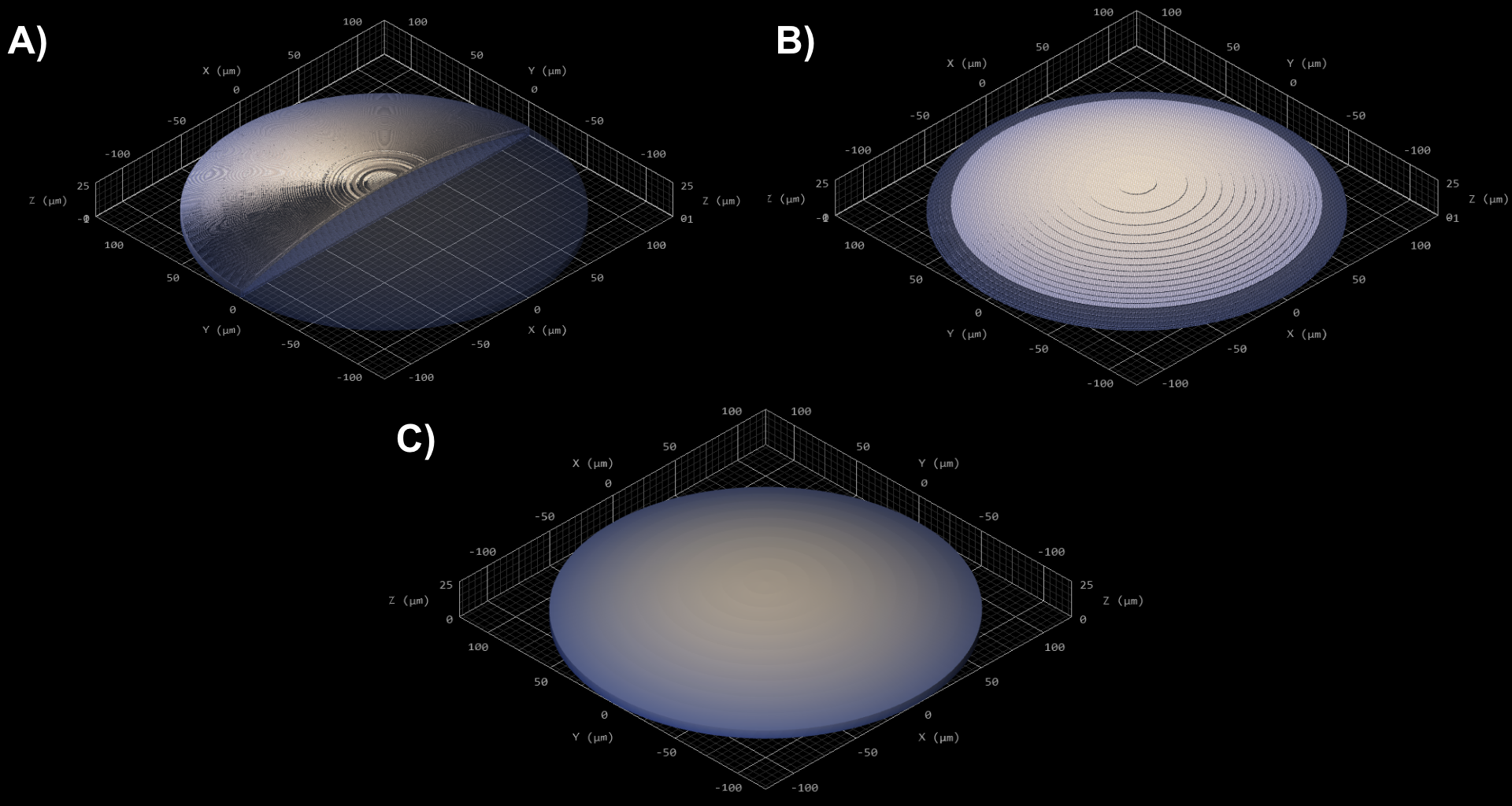}
    \caption{DeScribe import wizard models for 245\,micron lenses constructed using the shell and scaffold method. A) A simulation of the shell construction. B) A simulation of the scaffold construction. C) A simulation of the final lens resulting from combining the two components.}
    \label{fig:shell+scaff}
\end{figure}

An alternative to the solid printing process we first attempted is changing to a shell and scaffold print, where printing parameters such as laser power, slicing, hatching, and write speed can be set differently for the inner and outer regions of the lens. While the scaffold can take several shapes, of particular interest is the planar method, which allows for horizontal floors to provide all of the support to a lens. Ristok\cite{Ristok_2022} used a 2 component lens featuring a coarsely sliced core and a finely sliced shell to improve manufacturing speeds in printing single microlenses; the 2 component lens achieved similar results to one printed with only the fine shell parameters. Rather than printing 2 components, we used the built in planar slicing to mimic this approach; doing so also mimics some of the capabilities of the 2GL printing method for which IPX-Clear was originally produced. The scaffold (see \autoref{fig:shell+scaff}B) consisted of 1\,micron sliced planes with a 0.5\,micron hatch printed with a laser power of 17.5\,mW; these values were chosen because they are within the expected voxel, so the material should be fully polymerized via the 2PP method and we will not need to rely on UV curing to homogenize the interior. Over the scaffold was a more finely sliced and hatched shell (0.2\,microns) to provide the optical quality surface needed for the microlens array to be functional (\autoref{fig:shell+scaff}).

\begin{figure}
    \centering
    \includegraphics[width=0.95\linewidth]{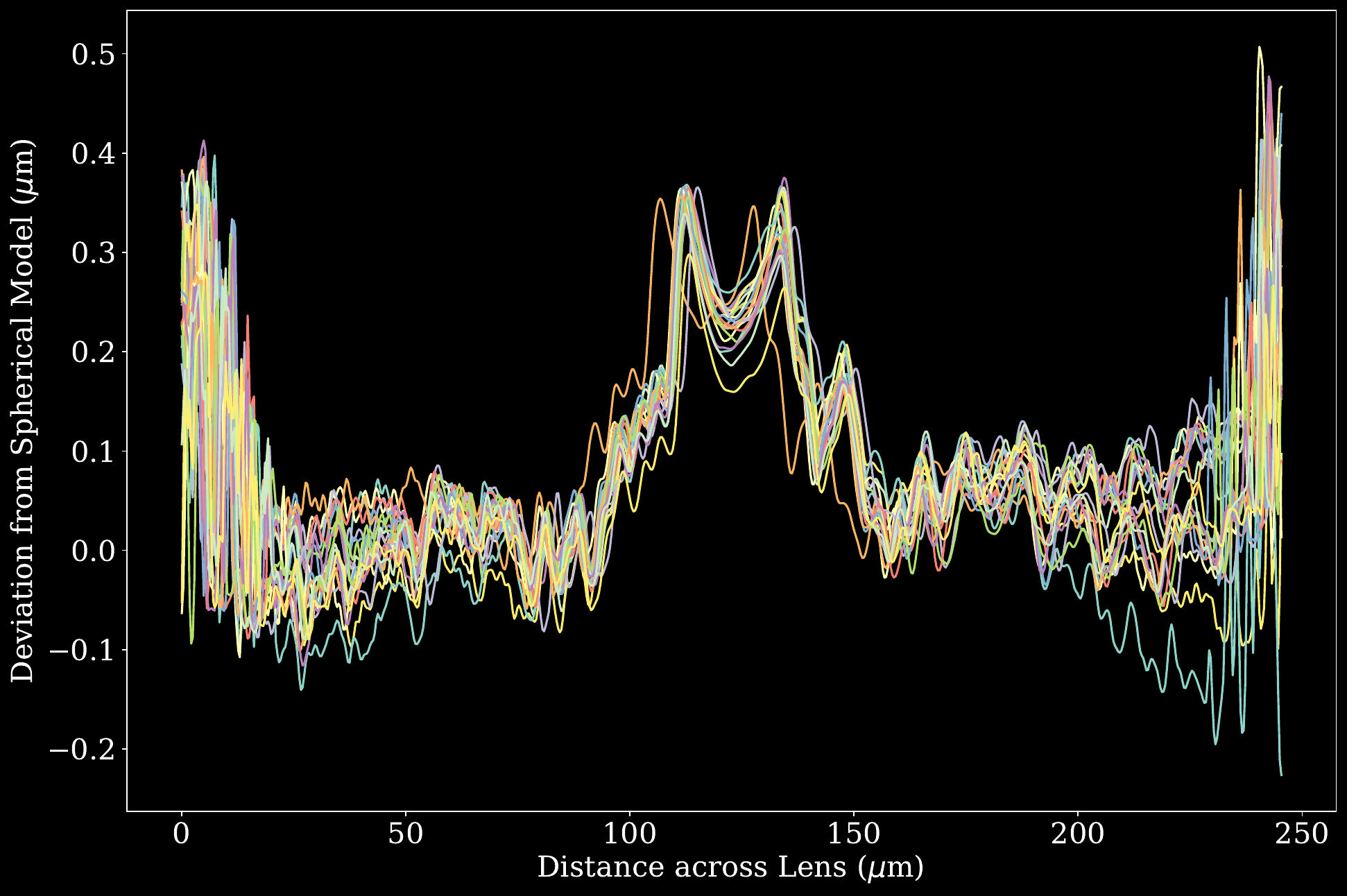}
    \caption{The deviation from a best-fitting spherical model of all lenses in the most recent 2x10 prototype array as a function of distance along a slice running parallel to the z-axis.}
    \label{fig:i51}
\end{figure}

This design allows for additional parameters in the form of shell thickness and differing laser powers for the outer shell as compared to the inner scaffold and the base. Having identified overdosing as an issue, particularly at the peak of the lens, the ability to decrease the laser power for the shell only allows for lower dosage at the peak without the height loss seen in the solid models at low laser power. \autoref{fig:i51} shows the profiles of a 2x10 lens prototype made using this methodology, with an average deviation from spherical across all lenses of $0.87\pm0.12$\,microns. Comparison of ellipsometry measurements between a solidly printed column and one printed with this methodology revealed no significant differences in refractive index, suggesting an internally homogeneous material. Further planned analysis will include examining cross sections of shell and scaffold lenses to ensure the same homogeneous interior seen in the solid lens cross sections.

\section{FUTURE WORK}
~~~Initial prototyping of the microlens array for LFAST has yielded promising results in the use of 2PP technology for optical components of astronomical telescopes and instruments. Work remains in refining the profile of the individual lenses which has been the major challenge in fabrication. We plan to implement changes to the profile of the design, using a iterative pre-compensation strategy to combat the uneven shrinkage\cite{lang_towards_2022} and reduce the deviation from spherical to $<0.5$\,microns at a minimum with an ideal case being a reduction to $<0.25$\,microns. We also plan to continue investigating the repeatability of this process for arrays of hundreds, and up to the total 2640 lenses required to combine all LFAST fibers into one common pseudo slit.
\begin{figure}
    \centering
    \includegraphics[width=0.95\linewidth]{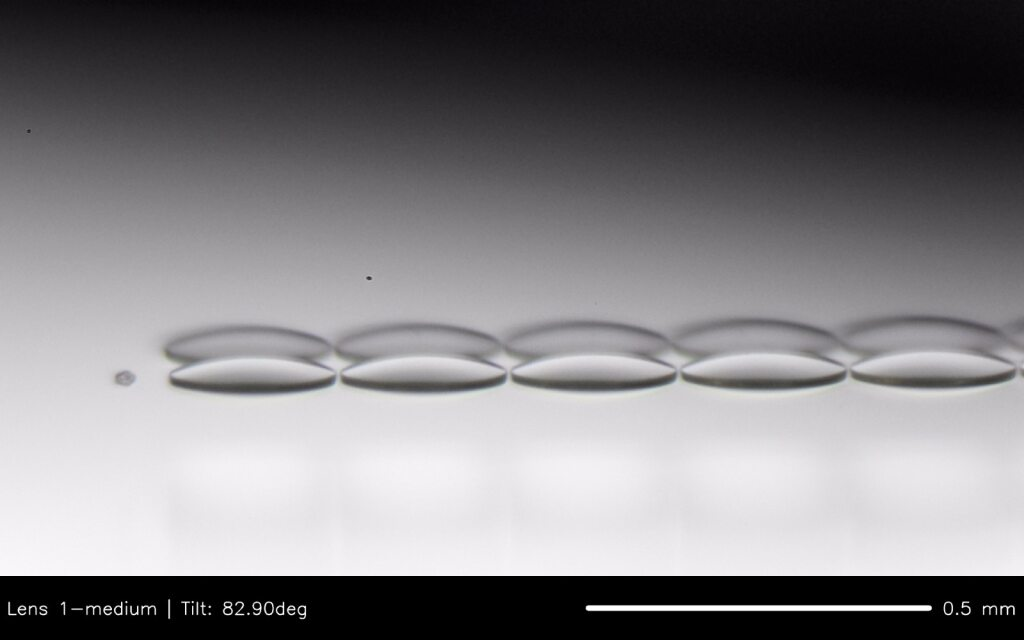}
    \caption{An optical microscopy image of a portion of the most recent 2x10 prototpye array. Images taken with a Microqubic MRCL700 3D Imager Pro.}
    \label{fig:placeholder}
\end{figure}

\acknowledgments 
 
We acknowledge funding support from the Penn State Materials Research Institute for this program. The LFAST project is supported by Schmidt Sciences LLC. The Center for Exoplanets and Habitable Worlds is supported by the Pennsylvania State University, the Eberly College of Science, and the Pennsylvania Space Grant Consortium. 

\bibliography{report} 
\bibliographystyle{spiebib} 

\end{document}